\documentclass{emulateapj}
\usepackage{graphicx}
\usepackage[dvips]{color}

\shorttitle{Chemical Signatures of the First Galaxies}
\shortauthors{Frebel \& Bromm}

\begin{document}
\title{Chemical Signatures of the First Galaxies: Criteria for
  One-Shot Enrichment}

\author{
Anna Frebel\altaffilmark{1} and 
Volker Bromm\altaffilmark{2}}

\altaffiltext{1}{Massachusetts Institute of Technology \& Kavli
  Institute for Astrophysics and Space Research, 77 Massachusetts Ave;
  Cambridge, MA 02139; afrebel@mit.edu}

\altaffiltext{2}{Department of Astronomy and Texas Cosmology Center,
University of Texas at Austin, 2511 Speedway, Austin, TX 78712-0259;
vbromm@astro.as.utexas.edu}

\begin{abstract}

We utilize metal-poor stars in the local, ultra-faint dwarf galaxies
(UFDs; $L_{\rm tot} \le 10^{5}\,L_{\odot}$) to empirically constrain
the formation process of the first galaxies. Since UFDs have much
simpler star formation histories than the halo of the Milky Way, their
stellar populations should preserve the fossil record of the first
supernova (SN) explosions in their long-lived, low-mass stars. Guided
by recent hydrodynamical simulations of first galaxy formation, we
develop a set of stellar abundance signatures that characterize the
nucleosynthetic history of such an early system if it was observed in
the present-day universe. Specifically, we argue that the first
galaxies are the product of chemical ``one-shot'' events, where only
one (long-lived) stellar generation forms after the first,
Population\,III, SN explosions. Our abundance criteria thus constrain
the strength of negative feedback effects inside the first
galaxies. We compare the stellar content of UFDs with these one-shot
criteria. Several systems (Ursa Major\,II, and also Coma Berenices,
Bootes\,I, Leo\,IV, Segue\,1) largely fulfill the requirements,
indicating that their high-redshift predecessors did experience strong
feedback effects that shut off star formation.  We term the study of
the entire stellar population of a dwarf galaxy for the purpose of
inferring details about the nature and origin of the first galaxies
``dwarf galaxy archaeology''. This will provide clues to the connection of
the first galaxies, the surviving, metal-poor dwarf galaxies, and the
building blocks of the Milky Way.

\end{abstract}


\section{Introduction}
One of the important unsolved problems in cosmology is understanding
the formation of galaxies (e.g., \citealt{benson10, mo10}).  To make
progress on this complex issue, it is advantageous to investigate
simple systems that allow us to study the basic processes that led to
their origin and evolution.  Dwarf galaxies are such objects and can
be studied both observationally and theoretically
\citep{mateo98}. They are generally old, metal-poor, have no gas and
thus no longer support star formation (\citealt{tolstoy_araa} for a
review). These conditions make them ideal candidates to constrain
theoretical models for star and galaxy formation in the early universe
\citep{bromm_araa11}. One particularly important question is to
elucidate the role of feedback processes in galaxy formation (see
\citealt{silk11}).  Again, early dwarf galaxies may provide us with an
ideal laboratory to test the physics of feedback (e.g.,
\citealt{dekel, ciardi,mashchenko08,maio10}).

In particular, the population of faint dwarf galaxies discovered in
the Sloan Digital Sky Survey (SDSS) that surround the Milky Way (MW)
offer a unique way to investigate these topics. Due to their simple
nature, these so-called ultra-faint dwarf galaxies (UFDs), here
defined to have $L_{\rm tot} \le 10^{5}\,L_{\odot}$
(\citealt{martin08}), are the closest local analogs to the first
galaxies. They are believed to have had only one or a few early star
formation events, but have been quiescent ever since (e.g.,
\citealt{koch_biermann}). Hence, they should retain signatures of the
earliest stages of chemical enrichment in their stellar populations.
Indeed, these systems are very metal-poor, and extend the
metallicity-luminosity relationship of the classical dwarfs down to
$L_{\rm tot} \sim 10^{3}\,L_{\odot}$ (see \citealt{kirby08} for more
details).  High-resolution spectroscopy \citep{ufs,norris10, leo4}
further showed that the abundances of faint dwarf galaxy stars
resemble those of similarly metal-poor Galactic halo stars. This
suggests that chemical evolution is universal, at least at the
earliest times which are probed by the most metal-poor, and thus
presumably the oldest, stars.
The same chemical trends have also been found in a few stars with
$-4.0\leq\mbox{[Fe/H]}\leq-3.5$ \citep{scl, tafelmeyer10} located in
the more luminous, classical dwarf spheroidals (dSph) Sculptor and
Fornax. However, at higher metallicity ($\mbox{[Fe/H]}>\sim-2.5$), the
stellar ([$\alpha$/Fe]) abundances of both systems deviate from those
of Galactic halo stars (e.g., \citealt{geisler05}), indicating a
different evolutionary timescale and multiple star formation
events leading to extensive metal-rich stellar components
\citep{tolstoy04}. This high level of complexity has been established
for all the classical dSphs down $L_{\rm tot} \sim 10^{5}\,L_{\odot}$,
making it difficult to directly connect them to the first galaxies.

The UFDs thus provide us with a tool for performing ``dwarf galaxy
archaeology''.  This terminology builds on the more general concept of
``stellar archaeology'' which posits that the chemical composition of
the early universe is preserved in the atmospheres of individual
metal-poor stars. Specifically, dwarf galaxy archaeology involves the {\it entire}
stellar content of a dwarf galaxy, including the more metal-rich
stars, in contrast to the focus on single metal-poor stars in the
traditional approach. With their relatively
limited number of stars, the least luminous galaxies are ideal
candidates for dwarf galaxy archaeology.  As opposed to the MW halo, which
was assembled through multiple merger and accretion events, the lowest
luminosity dwarfs likely did not form via extensive hierarchical
merging \citep{wise07, greif08, bovill09}. Their entire stellar
population, therefore, directly traces early star and galaxy
formation. The more metal-rich stars, with $\mbox{[Fe/H]}\gtrsim-2.0$,
in these faint metal-poor galaxies will provide the strongest
constraints on the star formation history, and hence the formation and
evolution of the host system. 

In particular, dwarf galaxy archaeology facilitates establishing the
connection between the surviving UFDs, the first galaxies and the
building blocks that formed the MW halo. This is important
since recent abundance studies have suggested early, accreted analogs of
today's UFDs to have played a significant role in building up the
metal-poor tail of the Galaxy \citep{ufs, leo4, norris10}. The Galactic
halo contains a significant number of extremely metal-poor stars
(with $\mbox{[Fe/H]}<-3.0$; e.g., \citealt{ARAA}), including some
objects with $\mbox{[Fe/H]}<-5.0$
(\citealt{HE0107_Nature,HE1327_Nature_short}). Their abundances have
been attributed to individual Population\,III (Pop\,III) SN yields
(e.g., \citealt{UmedaNomotoNature, tominaga07_b}) which provides a
key empirical diagnostic for Pop\,III nucleosynthesis and overall
constraints on the nature of these progenitors. Consequently, since
these stars are likely nearly as old as the universe, their 
origin may reside in small, early
systems. A better understanding of this connection is vital for
understanding chemical enrichment and star formation in the very early
universe \citep{karlsson11}.

In this paper, we specifically suggest that 
the stellar abundance record preserved in the metal-deficient dwarf
galaxies contain crucial hints on how effective early feedback effects
were in suppressing star formation. This endeavor
is complemented by the confluence of two recent
developments: The availability of large-scale parallel supercomputers
allowing ever more realistic simulations of early structure formation,
and increasingly detailed observations of stars in these UFDs.

\section[]{Cosmological Context}
The purpose of this section is to summarize those aspects of recent ab
initio simulations of first galaxy formation (see Bromm \& Yoshida
2011 for a review and further references) that provide us with the
theoretical underpinning and guidance in formulating potentially
observable chemical abundance signatures that may be found in a first
galaxy. We begin by discussing ideas on where the first stars and
galaxies form, and then turn to early metal enrichment.

\subsection{Early Star Formation Sites}
In a $\Lambda$CDM universe, structure formation proceeds
hierarchically, with small dark matter halos merging to form larger
ones. The first stars are expected to form in minihalos, collapsing
at $z\simeq 20 - 30$ \citep{tegmark97} and comprising
masses of $\sim 10^6$\,M$_{\odot}$. These minihalos host a small multiple
of predominantly massive Pop\,III stars
\citep{turk09,stacy10}. The individual masses of these first stars are
thought to be of order $\sim 100$\,M$_{\odot}$ \citep{abel_sci,
bromm02, yoshida08}, distributed according to a still uncertain
initial mass function (IMF).  It is likely, however, that a range of
masses towards lower and higher values would have been present
(e.g., \citealt{clark11, greif11}). Massive Pop\,III stars will exert strong
feedback on their host halos and the surrounding intergalactic medium
(IGM), through both radiative and SN feedback \citep{bromm03,
ciardi,alvarez06}, removing gas from the shallow potential well of the
minihalo, thereby quenching star formation.

A second round of star formation must have occurred in more massive
systems whose deeper potential wells were able to reassemble the
photo- and SN-heated gas from the diffuse IGM. It has been argued that
this can occur within so-called atomic cooling halos \citep{oh02},
having total masses of $\sim 10^8$\,M$_{\odot}$ and collapsing at
redshifts of $z\simeq 10-15$. Such systems have virial temperatures of
$T_{\rm vir}\simeq GM_h m_{\rm H}/(R_{\rm vir} k_{\rm B}) \sim 10^4$\,K,
where $M_h$ and $R_{\rm vir}$ are the halo mass and radius. At
these temperatures the gas can cool via excitation of atomic hydrogen
lines, without molecular hydrogen. Atomic cooling halos have been
proposed as the sites of the first bona-fide galaxies \citep{bromm09},
where a ``galaxy'' connotes a long-lived stellar system which can
sustain an interstellar medium, and extended, self-regulated, episodes
of star formation.

\subsection{Early Metal Production}

Assuming that they are plausible candidates for UFD progenitors,
hydrodynamical simulations of the formation of atomic cooling halos
prior to reionization are ideal for developing an understanding for
the nature of the building blocks and their connection to the first
galaxies, as well as any surviving dwarf galaxies. Since the
simulations are approaching the goal of ab initio calculations without
the need for recipes to model star formation and feedback effects
\citep{wise07,wise08,greif10}, the results are not affected by a
particular prescriptions for these processes.

Simulations indicate that an atomic cooling halo has of
order 10 progenitor minihalos \citep{wise07,greif08}. Each minihalo
in turn will harbor of order one SN explosion.  The latter prediction
is robust, and does not rely on a detailed knowledge of the Pop\,III
IMF. A minihalo will have a few $1000$\,M$_{\odot}$ of cold, dense gas
available for star formation \citep{yoshida03, yoshida06}. Assuming a
star formation efficiency of order 10 per cent, one has a few
$100$\,M$_{\odot}$ in stars.  For a top-heavy IMF, this would result
in of order one SN; for a normal, Salpeter-like IMF, one needs $\sim
100$\,M$_{\odot}$ of stellar mass to trigger one SN. In both
situations, we would have the same number of SNe per
minihalo. Consequently, a given Pop\,II star that formed in an atomic
cooling halo would be enriched by at most $\sim 10$ SNe, with an
element distribution that depends on the details of the turbulent
mixing of the metals.

One key simulation result is that the center of the emerging atomic
cooling halo is already enriched with heavy elements, to average
levels of $\sim 10^{-3} Z_{\odot}$ but with a spread of roughly
$\pm1$\,dex around this mean \citep{greif10, wise11}.  Overall, this
level of enrichment appears to be a robust expectation
\citep{johnson08}. Moreover, this spread could only be prevented in
cases where a strong Lyman-Werner radiation background is present, so
that H$_2$ can be destroyed inside the progenitor minihalos,
suppressing star formation and concomitant metal enrichment prior to
the collapse of the atomic cooling halo (e.g.,
\citealt{haiman97,johnson07}). However, this situation is thought to
be quite rare \citep{dijkstra08}. The typical atomic cooling halo will
therefore already be metal-enriched, and will eventually host Pop\,II
stellar systems. The enrichment history prior to the formation of
those second-generation stars is thus relatively simple, and solely
determined by massive SN yields.

Atomic cooling halos, at least at the low-mass end, may thus provide
environments for chemical ``one-shot'' events: their Pop\,II
starburst, synchronized to within roughly the dynamical time of the
central gas cloud of a few $10^5$\,yr \citep{greif08} might be able to
drive any remaining gas out of their shallow potential wells. This
conjecture needs to be tested with forthcoming highly-resolved
simulations of the central starburst. To foresee the outcome, we
consider the following approximate arguments: Just prior to the onset
of the initial starburst, of order $10^5 M_{\odot}$ of cold, dense gas
would have assembled. Again assuming a star formation efficiency of 10
per cent on these scales, we expect a star cluster of total mass $\sim
10^4 M_{\odot}$ to form. Such central clusters would have
luminosities of $10^3 - 10^4 L_{\odot}$, similar to the total
stellar luminosity observed in UFDs (e.g., \citealt{martin08}).
For a standard IMF, the starburst would be
accompanied by $\sim 100$ core-collapse SNe with an explosion energy
of $\sim 10^{51}$\,erg each. The total SN energy would then be
comparable to the gravitational binding energy of an atomic cooling
halo at $z\simeq 10$ \citep{mackey03}, rendering a complete removal of
all remaining gas at least plausible. A second feedback effect that
will act to evacuate the post-starburst halo is heating due to
photoionization \citep{johnson09}. 

Such simple, postulated one-shot enrichment systems are the ``Rosetta
Stone'' of cosmic chemical evolution. If still observable, they would
be ideal objects for carrying out dwarf galaxy archaeology. Their surviving
Pop\,II stars would preserve the yields from the initial Pop\,III SNe
that had occurred in the progenitor minihalos without any subsequent
enrichment from events that operate on timescales longer than the
short dynamical time, such as type\,Ia SNe or asymptotic giant branch
(AGB) winds. A possible caveat that could act to mask the Pop\,III SN
yields is pollution of these ancient stars with accreted interstellar
material. However, such contribution is likely extremely small, and
can therefore be neglected (e.g., \citealt{poll}).

\section{Criteria for one-shot enrichment} 

Assuming the one-shot enrichment scenario, we now discuss what kinds
of chemical signatures might occur in a first galaxy. We highlight
specific abundance predictions throughout this section.

According to the simulations (see Section~2), any first galaxy is
expected to have been chemically enriched by one or a few SNe, but no
more than $\sim10$, corresponding to the number of precursor
minihalos.  To bracket the uncertainties in the primordial IMF, we
consider two SN types occurring during the assembly of a first galaxy:
conventional core-collapse SNe in the mass range of
10-140\,M$_{\odot}$ and pair-instability supernovae (PISNe) occurring
between 140-260\,M$_{\odot}$. Since PISNe are assumed to be rare, we
expect that no more than one minihalo hosted such an explosion,
whereas all other events were core-collapse SNe. Such a distribution
can be regarded as an example of early chemical enrichment, but
different proportions of the two SNe types are of course possible. For
simplicity, we do not consider such cases here. We note, however, that
any stars with $>260$\,M$_{\odot}$, while perhaps present in
minihalos, would directly collapse into black holes and thus not
contribute to the enrichment. The bulk of the metals is contained in
$\sim 10^{5}$\,M$_{\odot}$ of gas, the typical mass of a star forming
cloud in an atomic cooling halo \citep{greif10}.

\begin{figure*} 
\begin{center}
 \includegraphics[clip=true,width=17.5cm,bbllx=15,bblly=225,bburx=520,bbury=610]{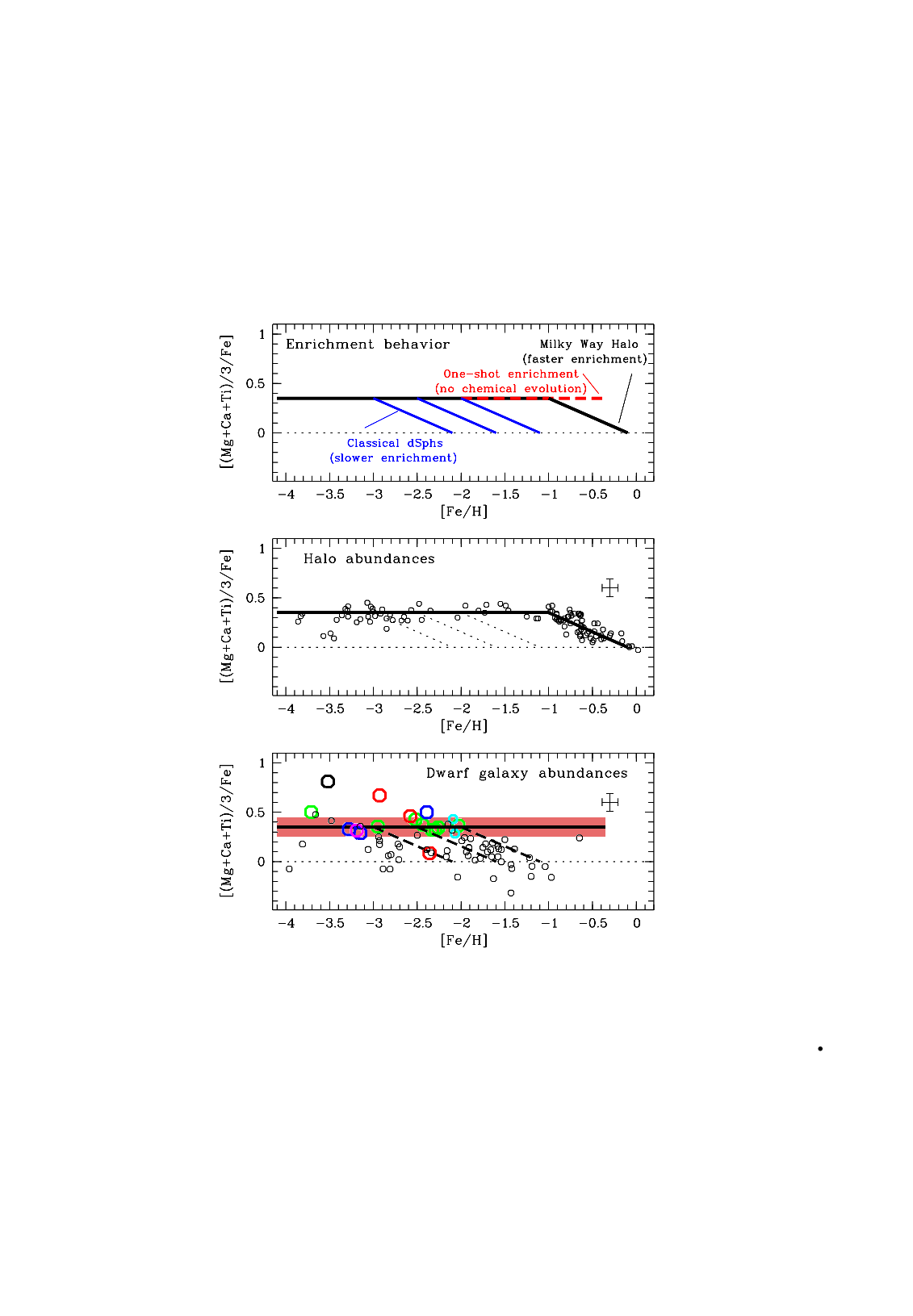}
 \caption{Top: Schematic representation of chemical enrichment in
   $\mbox{[$\alpha$/Fe]}$ vs [Fe/H] plane. The behavior for the Milky
     Way and dSph galaxies are shown together, with our prediction for
     a one-shot enrichment with no subsequent chemical evolution. The
     schematic behavior for the Milky Way and the dSphs has been
     deduced from the metal-poor data in the two lower panels.  A
     combined Mg-Ca-Ti abundance is used to represent the $\alpha$
     abundance typical for core-collapse SNe enrichment.
The dotted line indicates the solar ratio.
Middle: High-resolution $\alpha$-abundances of metal-poor stars from
\citet{cayrel2004} (halo) and \citet{fulbright} (thin/thick
disk). The diagonal dotted lines indicates the enrichment behavior of
the dSph galaxies (see bottom panel), which differs from that of the
Milky Way. A representative uncertainty is shown.
Bottom: High-resolution $\alpha$-abundances of metal-poor stars in the
classical dSph (small open black circles and several evolutionary path
are indicated with dashed lines;
\citealt{shetrone01,shetrone03,fulbright_rich,geisler05,
  aoki09,cohen09,scl,tafelmeyer10}) and UFD galaxies
(\citealt{feltzing09,ufs,norris10, leo4, norris10_seg}). Different
colors denote different UFD galaxies. Open red circles: Coma
Berenices, blue: Ursa Major\,II, pink: Leo\,IV, cyan: Hercules, green:
Bootes\,I, black: Segue\,1. The pink shaded region around
$\mbox{[$\alpha$/Fe]}=0.35$ depicts the predicted one-shot enrichment
behavior (with 0.1\,dex observational uncertainty) as set by the
$\alpha$-element enrichment caused by core-collapse SNe only.  }
\end{center}
\end{figure*}

\begin{figure*}
\begin{center}
 \includegraphics[clip=true,width=17.5cm,bbllx=15,bblly=325,bburx=550,bbury=530]{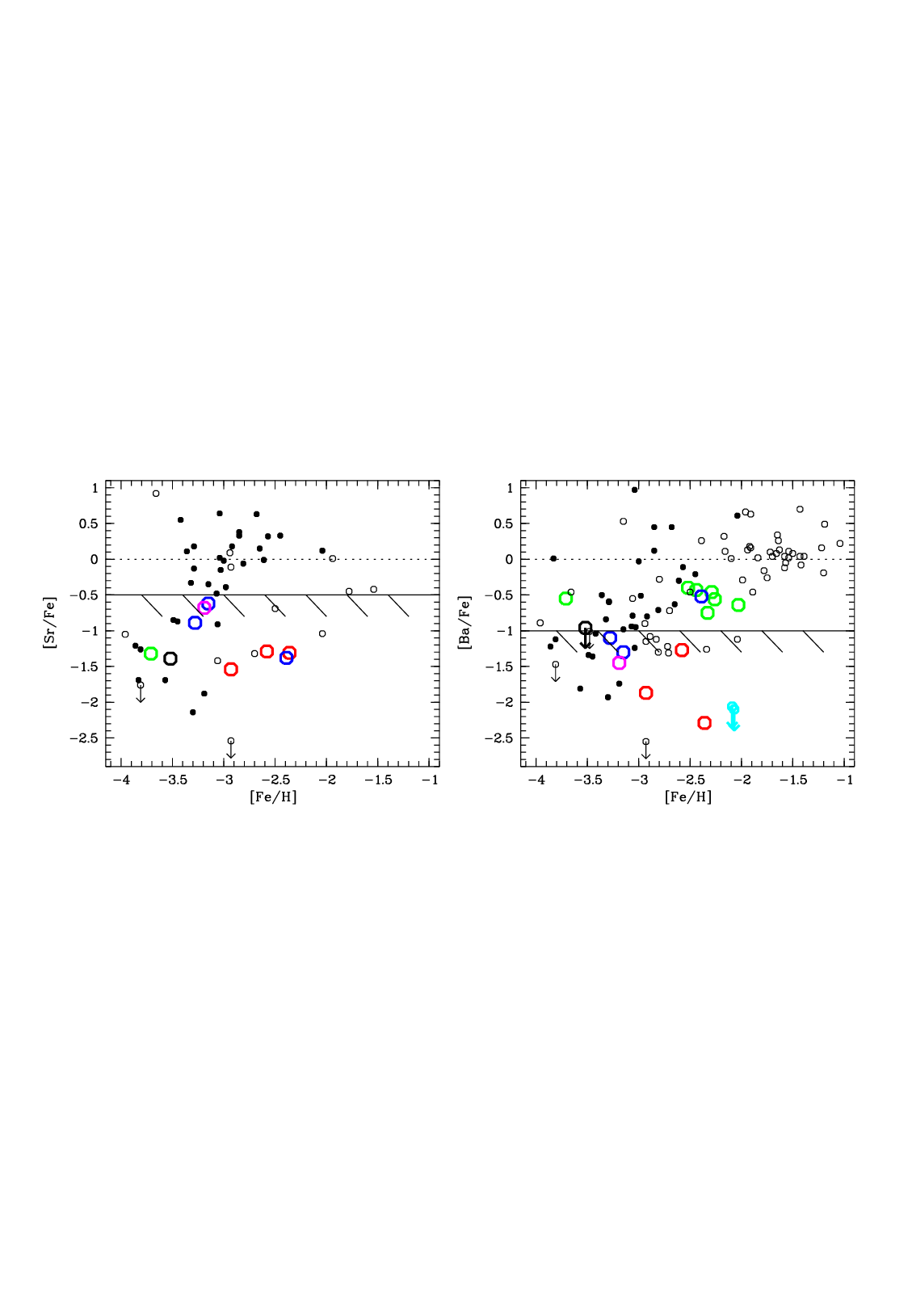}
 \caption{High-resolution neutron-capture abundance ratios [Sr/Fe] and
   [Ba/Fe] of UFD galaxy stars as a function of their metallicities
   [Fe/H], in comparison with halo and dSph galaxy stars.  Robust
   upper limits for a potential r-process enhancement are indicated by
   the solid horizontal lines in each panel (see text for discussion).
   Large open circles show different UFD galaxies; see Figure~1 for a
   description. Filled black dots represent halo stars
   \citep{cayrel2004, francois07}. Small open black circles show the
   abundances of stars in the classical dSph galaxies. }
\end{center}
\end{figure*}

We now consider the likely chemical signature of the
second-generation, low-mass metal-poor stars in the first
galaxy. Since these objects would be long-lived and thus still be
observable today, they provide a fossil diagnostic of these early
systems.  \textit{According to the one-shot scenario, no subsequent
  SNe would have contributed to the chemical inventory after the
  formation of the next (i.e. second) generation of stars.} This second
  generation included the first low-mass stars whose atmospheric
  abundances should thus preserve the chemical signatures of the
  Pop\,III progenitors.

Typical core-collapse SNe produce $\sim$0.1\,M$_{\odot}$ of Fe
\citep{heger_woosley10}. Considering the canonical example of diluting
this mass of Fe into a hydrogen gas of $10^{5}$\,M$_{\odot}$ leads to
a next-generation with a metallicity of $\mbox{[Fe/H]}=-3.25$. A
maximum of $\mbox{[Fe/H]}=-2.25$ would then be present as a result of ten
such SNe, and should reflect the average metallicity of the system.

Large abundance spreads in [X/H] are expected due to incomplete mixing
on the short dynamical timescale within the center of the first galaxy
\citep{greif10}. The second-generation starburst will occur on roughly
this timescale, so that metallicity inhomogeneities in the gas will be
reflected in the respective stellar abundances.  We note that any
inhomogeneous mixing occurring in a system would primarily affect
element ratios containing H, i.e., [X/H], but to a much lesser degree
those with two heavy elements, i.e., [X/Fe], assuming that no
differential mixing takes place on length and time scales relevant for
star formation. Hence, any spread for example in [Fe/H] is possible,
but little or no scatter in, e.g., [$\alpha$/Fe].  \textit{Consequently, large
variations of $\mbox{$\Delta$[Fe/H]}\sim1$\,dex or more around the
average systemic metallicities of $\mbox{[Fe/H]}\sim-2.3$ are expected
to occur.}

The core-collapse ejecta (e.g., \citealt{heger_woosley10}) have well
correlated Fe and $\alpha$-abundances (Mg, Ca, Ti, Si), resulting in
the characteristic metal-poor halo star signature of
$\mbox{[$\alpha$/Fe]}\sim0.35$ \citep{cayrel2004}. This abundance
level is shown in Figure~1 in the middle panel, and reproduced in the
top and bottom panels. In contradistinction, the bottom panel shows
abundances of stars in the classical dwarf galaxies whose chemical
enrichment proceeded on a slower timescale. Hence, stars with
$\alpha$-abundances below the halo value are found at lower
metallicities than $\mbox{[Fe/H]}\sim-1.0$. Again, this behavior is
reproduced in the top panel of the figure. We also show the currently
available abundance data of stars in the UFDs, which we discuss below
and in Section~4.

The halo and dwarf galaxy stars thus clearly show the standard
``multi-shot'' enrichment histories enabled by many generations of
SNe, i.e. extended phase(s) of chemical evolution and star formation.
Such normal chemical evolution is reflective of feedback processes
having a much less severe or even no impact on the existing conditions
of the host galaxy.

Following the one-shot scenario with enrichment by massive Pop\,III
stars, for the first galaxies, we therefore predict average values of
[$\alpha$/Fe] as set by the sample of \citet{cayrel2004} with a range
of $\pm0.1$\,dex (approximately the measured standard
deviation). \textit{A first galaxy should thus contain no stars, not
  even at higher metallicity ($\mbox{[Fe/H]}>\sim-2.0$), that show
  $\alpha$-abundance ratios systematically less than the halo value or
  even with the solar ratio, $\mbox{[$\alpha$/Fe]}=0.0$. This behavior
  is illustrated in Figure~1 (top and bottom panels).}
Such low values would indicate star formation after any of the more
massive Pop\,II stars eventually exploded as SNe\,Ia, adding iron to
the galaxy, in contradiction to the one-shot assumption.

The top panel in Figure~1 thus summarizes the basic three possible
enrichment histories for a given galaxy.

Late-time AGB or SN\,Ia enrichment from lower mass stars would
eventually occur in a one-shot system, but only after the initial
Pop\,II starburst, and after the remaining gas was blown out of the
system (see the discussion in Section~2). \textit{Consequently, there
  should not be any signs of general s-process enrichment by AGB stars
  despite the fact that some of these Pop\,II stars with intermediate
  masses must have gone through an AGB phase to later provide
  s-process material as well as carbon.}  The only exceptions would be
individual metal-poor stars with strong s-process (and carbon)
enhancements due to a mass transfer across a binary system.

If an r-process occurred in one of the core-collapse SNe (e.g., one
with a 10-20\,M$_{\odot}$ progenitor), small amounts of
neutron-capture material would be produced. Unfortunately, no
theoretical r-process yields are available for direct comparison with
observations. The one exception is the ``weak'' r-process
investigation by \citet{izutani}, focusing only on the production of
the light neutron-capture elements Sr, Y, and Zr. For two different
progenitor masses and ``normal'' explosion energy of $E_{51}\sim1$,
their models yield Sr ejecta of $M(\rm
Sr)\sim1\times10^{-8}$\,M$_{\odot}$ (their 13\,M$_{\odot}$ model) and
$M(\rm Sr)\sim1\times10^{-7}$\,M$_{\odot}$ (25\,M$_{\odot}$ model),
with respective Fe yields of $M(\rm
Fe)\sim6\times10^{-2}$\,M$_{\odot}$ and $M(\rm
Fe)\sim2\times10^{-1}$\,M$_{\odot}$. Diluting these yields in
10$^{5}$\,M$_{\odot}$ of H gas, yields low values of
$\mbox{[Sr/H]}\sim-5.8$ and $-4.8$ depending on the model. The
corresponding [Sr/Fe] values are $\sim-2.3$ and $-1.9$, with
corresponding [Fe/H] values of $-3.5$ and $-2.9$. Their higher
explosion energy 25\,M$_{\odot}$ model produces more Sr (which they
conclude to be the appropriate progenitors for their group of ``weak''
r-process stars), although the results appear to be very sensitive to
model parameters. Considering just their ``normal''-energy SN yields,
very low levels of [Sr/Fe] must have been present in a first galaxy.
By extension, [Ba/Fe] values must have been even lower, assuming the
ratios of $\mbox{[Sr/Ba]}\sim0.4$ of typical stars in the halo sample
\citep{francois07}.

We arrive at a more general, heuristic limit on the r-process
contribution in a first galaxy, specifically on the Sr and Ba
abundances, as follows: In the MW, the s-process is known to dominate
the chemical evolution of neutron-capture elements above
$\mbox{[Fe/H]}>-2.6$ \citep{simmerer2004}, as provided by AGB
stars. 
We use metal-poor stars from the literature to estimate the general
trends of [Sr/Fe] and [Ba/Fe] in stars with $-2.6<\mbox{[Fe/H]}<-1.7$.
These stars characterize early gas clouds that were significantly
enriched in s-process elements.
\textit{Consequently, stars with abundances lower than the general
  trend can be regarded to represent either an early enrichment by the
  very first individual AGB stellar winds or an r-process enrichment as
  provided by core-collaps SNe, in same way as for stars at lower
  metallicities before the onset of any AGB enrichment. }
These lower-than-average values are $\mbox{[Sr/Fe]}<-0.5$ and
$\mbox{[Ba/Fe]}<-1.0$, as indicated in Figure~2. Also shown are the
abundances of halo, classical dwarf galaxy and UFD stars. These
empirical limits are somewhat higher than the values derived from the
\citet{izutani} calculations, but they represent a more robust upper
limit to a pure r-process enrichment by core-collapse SNe.

We now consider the yields of a PISN, in addition to core-collapse SN
enrichment, and how they would change the chemical make-up of a first
galaxy.  There are two main chemical signatures expected to be found
in metal-poor stars if a PISN had enriched the system as well.  Since
there are 10 independent star forming halos (the $\sim10$ progenitor
minihalos that merge to form the first galaxy), it is very likely that
the PISN will be accompanied by up to $\sim10$ core-collapse SNe.
\textit{Consequently, the previously described enrichment pattern
  would be present, but additionally, individual stars would display a
  more or less clean PISN signature.} The strength of the PISN
signature in a given star will depend on the details of the
stochastic, inhomogeneous mixing of the ejecta from all contributing
SNe \citep{karlsson08}.  Due to nearly constant yields of C, Mg, and
Ca independent of the progenitor mass \citep{heger2002}, respective
next-generation stellar abundances would be high, e.g.,
$\mbox{[Ca/H]}\sim-2.0$, albeit with a spread of $\pm0.5$\,dex.  Fe,
on the other hand, with almost arbitrary yields, is not directly
predictable. With increasing progenitor mass, vast quantities of Fe,
and also O, are produced, up to several tens of solar masses (see
fig.~1 in \citealt{heger2002}). Consequently, such a star could have
$\mbox{[Fe/H]}\sim-3.0$ or even less, or $\mbox{[Fe/H]}\sim-1.0$ or
more. Either way, the [Mg/Fe] and [Ca/Fe] ratios would likely differ
from that of typical halo stars. \textit{Another characteristic
  signature of PISN-enriched gas is a complete lack of neutron-capture
  elements. Hence, any star formed from such material would have no
  detectable neutron-capture elements.}

\section{The Dwarf Galaxy Archaeological Record} 

In Figures~1 and 2, we have summarized the currently available
abundances obtained from high-resolution spectroscopy of stars in
several UFDs as well as halo stars. We show [$\alpha$/Fe] as a tracer
of core-collapse enrichment (Figure~1, top), and the neutron-capture
elements [Sr/Fe] and [Ba/Fe] as a gauge for pure r-process material
(Figure~2). We have presented our predictions for the chemical
signatures in a first galaxy, following the one-shot enrichment of the
system as part of it assembly from Pop\,III hosting minihalos. For
convenience, we briefly summarize these predictions again:

\begin{itemize}

\item Only SNe from the first generation of stars
  would contribute to the chemical inventory of a first galaxy.

\item Inhomogeneous mixing would lead to large variations of
  $\mbox{$\Delta$[Fe/H]}\sim1$\,dex or more around the average
  systemic metallicities.

\item A first galaxy should contain no stars, not even at higher
  metallicity ($\mbox{[Fe/H]}>\sim-2.0$), that show $\alpha$-abundance
  ratios systematically less than the Galactic halo abundance value of
  $\mbox{[$\alpha$/Fe]}\sim0.35$.

\item No general signature of s-process (+ carbon) enrichment from AGB
  stars should be identifiable in the surface abundances of long-lived
  low-mass stars in a first galaxy.

\item If PISN also occurred in a first galaxy, then in addition to the
  previous points, individual stars would display a PISN signature,
  i.e. high $\mbox{[$\alpha$/Fe]}$ values very different from that of
  halo stars.

\item A complete lack of neutron-capture element enrichment by PISN
  would lead to next-generation stars with no detectable neutron-capture
  elements.

\end{itemize}

Based on these predictions, we can now asses whether any of the
surviving UFDs chemically resemble a one-shot enrichment first galaxy.

We begin with enrichment by core-collapse SNe. To what extent do the
UFD stars show an [$\alpha$/Fe]-enhancement of $\sim0.35$\,dex,
especially at metallicities of $\mbox{[Fe/H]}>-2.5$?  This signature
has indeed been found for most stars in the UFDs that have detailed
chemical abundances available, and it has been suggested that the
observed elements originated from canonical core-collapse events in
the same way as found for halo stars (e.g., \citealt{ufs,
  leo4}). Unfortunately, not many metal-rich stars are present and
also observable in these systems, resulting in no data at
$\mbox{[Fe/H]}>-2.0$.

Bearing in mind that only one calculation of r-process yields is
currently available, the prediction of low neutron-capture abundances
is in reasonable agreement with the overall depleted abundances of Sr
and Ba (e.g., many stars have $\mbox{[Ba/Fe]}<-1.0$) in all of the
UFDs. We also note that no metal-poor star with s-process enhancement
from a binary mass transfer has yet been conclusively identified in
any of the known UFDs.\footnote{\citet{ufs} found a star with apparent
radial velocity variations (two measurements only) which is possibly
s-process enhanced. A clarification, however, would require measurements of additional
neutron-capture elements.}
Recent medium-resolution spectroscopic studies \citep{kirby08,
norris_boo,norris10_booseg} showed that all of the UFDs have large
[Fe/H] spreads of $\sim1$\,dex or more, and reaching below
$\mbox{[Fe/H]}=-3.0$. Moreover, some have average metallicities as low
as $\mbox{[Fe/H]}\sim-2.6$ (Leo\,IV, Hercules), which is less than
that of the most metal-poor globular clusters. None of the systems
with $L_{\rm tot} \lesssim 10^{5}\,L_{\odot}$ have averages of
$\mbox{[Fe/H]}>-2.0$. These low metallicities agree well with the
estimates for Fe enrichment from up to 10 core-collapse SNe in a
first galaxy. The large abundance spread is also in agreement with
simulation results, reflecting inefficient mixing.

Out of the six UFDs with available abundance data, Ursa Major\,II
(blue) best fulfills the criteria for being a candidate first galaxy
fossil.  The one star with a higher Ba abundance is possibly an
externally enriched binary star, so no strong conclusions can
currently be derived from this object.  Coma Berenices (red) and
Bootes\,I (green) are good candidates as well, although the Bootes\,I
stars have higher Ba abundances than our limit, and Coma Berenices'
highest metallicity star shows a decreased [$\alpha$/Fe]. However,
future r-process predictions will reveal whether this is necessarily
inconsistent with a one-shot enrichment. Intriguingly, Hercules (cyan)
appears to be different. Not shown in Figure~1 (bottom panel) are the
Ca abundances of several stars \citep{aden11}, which show large
variations ranging from subsolar to $\mbox{[Ca/Fe]}\sim0.3$. We
tentatively rule out Hercules as a candidate, along the same line as
the more luminous classical dwarfs. We note, however, that
\citet{aden11} had one star in common with \citet{koch_her}, but
derived a 0.4\,dex lower Ca abundance, which is their lowest value in
the sample. This is somewhat puzzling, but given that the Ca spread is
about twice the discrepancy, it seems reasonable to assume that this
galaxy indeed has a significant abundance spread in this element.
Leo\,IV (pink) and Segue\,1 (black) at present contain too few data to
arrive at a meaningful conclusion.  Hence, additional high-resolution
abundance studies of more stars in each of these as well as other
systems are required.  But new Segue\,1 results (A. Frebel et
al. 2012, in prep.) already indicate this system to be in agreement
with a pure core-collapse SN enrichment.  Only with more metal-rich
stars can it be revealed whether the one-shot conjecture holds or if
evidence for extended star formation and chemical evolution can be
found. Either result would provide important constraints on early
feedback processes.

{\it Are there any hints for a potential PISN enrichment?}  Since the
faintest galaxies have very low average metallicities and truncated
star formation, UFDs provide the perhaps best chance to ever detect
the chemical signature of a PISN event.  Identifying the PISN
signature is, however, difficult given that the predicted very high yields 
may lead to stars with much higher metallicity compared with a regular
core-collapse enrichment (see Karlsson et al. 2008).
Even if a first galaxy did not survive until the present time, it is
likely that individual stars that formed in these environments passed
into larger systems through merger events, possibly into surviving
galaxies, e.g., Hercules with $\log(L/L_{\odot})=4.6$, or more luminous
systems such as Draco and Sculptor. Hence, individual stars in
systems more luminous than the faintest UFDs with extensive star
formation and chemical evolution could still preserve the rare signature
of PISNe. 

One interesting galaxy in this context is Hercules, for which
\citet{koch_her} measured the abundances of many elements of two
member stars. Both Ba limits are $\mbox{[Ba/Fe]}<-2.1$, and among the
lowest values ever measured.\footnote{It should be noted that these
stars with $\mbox{[Ba/H]}<-4.15$ do not have the lowest [Ba/H] values
or limits. Many halo stars have abundances $\mbox{[Ba/H]}\le-5.0$
($\sim10$ stars found in the compilation of \citealt{frebel10}; e.g.,
from \citealt{francois07,lai2008}).} While the average metallicity of
Hercules is $\mbox{[Fe/H]}\sim-2.6$ (but with a spread of more than
$\sim1$\,dex in Fe; \citealt{kirby08, aden11}), these two stars have
rather high metallicities of $\mbox{[Fe/H]}\sim-2.0$. Both stars have
high Mg/Fe ($\sim0.8$) and low Ca/Fe ratios ($\sim0.05$), but their
Ti/Fe corresponds to the typical halo value.
\citet{koch_her} speculated that in order to produce large Mg/Ca
ratios in a ``next-generation'' star (i.e., the observed star),
Hercules would have to have been enriched by fewer than 11
core-collapse SN events, consistent with our picture of first galaxy
enrichment. Considering the above PISN enrichment criteria, Hercules
could be a candidate site for a PISN pre-enrichment, and for testing
the predictions for PISN yields \citep{karlsson08}. However, the
recent low stellar Ca/Fe abundances of \citet{aden11} in Hercules
complicate the situation, and indicate that this system is not a 
candidate first galaxy. Additional observations will be helpful to
fully understand the chemical evolution of this galaxy.

Another interesting case is a star with $\mbox{[Fe/H]}\sim-3.0$,
located in the classical dSph Draco, and having an upper limit of
$\mbox{[Ba/Fe]}<-2.6$ \citep{fulbright_rich}.  Strontium is similarly
depleted and no other neutron-capture elements could be detected.  On
the contrary, all other, more metal-rich stars in Draco do not show
this behavior \citep{cohen09}. This star is thus highly unusual, but
similar to the stars in Hercules, except for its lower
metallicity. This difference might be due to the arbitrary Fe yields
of PISNe, inhomogeneous mixing or simply the gas mass available for
mixing, which could be $\sim10$ times more than in a Hercules-like
object. \citet{fulbright_rich} found that no ordinary SN model could
account for the lack of neutron-capture elements in this star, and its
origin is still uncertain.  Given that this galaxy is only $\sim$10
times more luminous than Hercules, and thus lies on the low-luminosity
tail of the classical dSphs, it could plausibly have assembled from
several first galaxies. Hence, some individual stars could have
preserved their PISN signature throughout their cosmic merger journey
while showing signs of extended star formation at the same time.

\section{Constraints on the Nature of First Galaxies}

Next to atomic cooling halos, minihalos have been suggested as UFD
progenitors (e.g., \citealt{salvadori09, bovill09}). The minihalo
environment may, however, face a problem, at least for the lowest-mass
minihalos that are close to the threshold mass ($\sim 10^6 M_{\odot}$)
required for H$_2$ cooling to become effective. In such minihalos the
available gas mass is only $10^{3} -10^{4}$\,M$_{\odot}$ (e.g.,
\citealt{yoshida06}), much lower than in the more massive atomic
halos. Consequently, any SN yield is much less diluted, generally
resulting in stars with higher metallicity than those in atomic
cooling halos.  Diluting 0.1\,M$_{\odot}$ of Fe into the available gas
mass yields a next-generation star with $\mbox{[Fe/H]}\sim-1.2$ (for
$10^{3}$\,M$_{\odot}$). Even the assumption that inhomogeneous mixing
would be able to produce a spread of $\pm1$\,dex around this value
could not explain extremely metal-poor stars with $\mbox{[Fe/H]}<-3$,
including the most metal-poor star in any UFD, BooI-1137 with
$\mbox{[Fe/H]}\sim-3.7$ \citep{norris10}. The only possibility would
be to limit the maximum Fe SN yield to $\sim0.001$\,M$_{\odot}$ (or
0.01\,M$_{\odot}$ for the larger dilution mass of
$10^{4}$\,M$_{\odot}$) in any minihalo in the early universe. Atomic
cooling halos with their larger gas reservoirs thus appear, at least
broadly, to be able to account for the existence of the
lowest-metallicity stars.

We point out that the minihalos invoked as UFD progenitors
\citep{salvadori09,bovill09} typically lie at the high mass end of the
minihalo range, thus largely circumventing this mixing problem as
well. Within the minihalo scenario, the same system would have to
first lead to the explosion of Pop\,III SNe, subsequently reassemble
the enriched gas inside its shallow potential well, and finally
trigger a second generation of star formation. For the atomic cooling
halo pathway, on the other hand, the sites for first and second
generation star formation are decoupled, thus alleviating the problem
of admitting Pop\,III pre-enrichment. Altogether, we thus favor the
atomic cooling halo path that can more readily explain the presence of
metal-poor stars. We stress, however, that there will be a continuum
of more complex enrichment histories, where multiple SN generations
and contributions from low-mass stars, corresponding to host systems
of subsequently larger mass, and therefore deeper potential wells (for
an alternative view, see \citealt{strigari08}). Observationally, this
sequence of cosmological formation sites corresponds to the progression
from the lowest-luminosity dwarfs, to classical dwarf spheroidals, and to
Magellanic-cloud type irregulars.

Assuming that the UFDs are chemical one-shot events, the observed
spread in Fe (i.e., [Fe/H]) suggests that mixing in these early
systems was inefficient.  Otherwise all stars would have nearly
identical abundances, similar to what is found in globular clusters
(e.g., \citealt{gratton04}).  We can thus infer that mixing in the
very first galaxies was largely incomplete, whereas globular clusters
must have formed in much different environments where turbulent mixing
would have been much more efficient. We repeat  that such
inefficiency would not yield a scatter in the elemental abundance
ratios [X/Fe], unless differential mixing among different elements
played an important role.

Hercules, having the highest luminosity of the examined
systems, exhibits an abundance behavior, e.g., a large spread in Ca, that
is suggesting
possible SN\,Ia enrichment. We therefore derive an upper limit to the
luminosity for candidate first galaxies of
$\log(L/L_{\odot})\sim4.5$. Bootes\,I, with a similar luminosity will
be an interesting object in this regard, and additional observations
will show if this system remains a good candidate. It is interesting
to note that Coma Berenices has recently been shown to be
a stable UFD with no signs of tidal stripping \citep{munoz10}, while
Ursa Major\,II appears to currently undergo disruption.
As for Leo\,IV, Simon et al. (2010) suggested that its entire
Fe content could have been provided by a single SN. If confirmed with additional
observations, the case of Leo\,IV would show that one-shot events do
take place, and that such simplistic galaxies, like the first
galaxies, can survive to the present day.

\section{Discussion and Conclusions}

The currently known UFD abundance record leads us, with the caveats
and qualifications discussed above, to derive the following
conclusions.  Independent of the question of whether UFDs are
surviving minihalos or atomic cooling halos, we suggest that at least
some of today's UFDs (Ursa Major\,II, Leo\,IV and possibly also Coma
Berenices, Bootes\,I) are likely to have been the results of chemical
one-shot events that occurred in the early universe. Given that atomic
cooling halos seem to be the more favorable environments for producing
low-metallicity stars that resemble the observed stellar populations
of the UFDs, these systems are plausible formation sites for the least
luminous galaxies.

As additional chemical abundances of individual dwarf galaxy stars are
measured, abundance gradient studies of the UFD galaxies will further
constrain the mixing efficiency. Stronger gravitational fields in the
center of a system would drive more turbulence that in turn would
induce mixing.  To properly interpret the data, in particular for
future observations with extremely large telescopes, the expected
signature of clustered star formation in the first galaxies needs to
be taken into account \citep{blandhawthorn10}. Since the UFDs are
ideal testbeds for various feedback processes, it will also be
interesting to study the carbon enrichment and the spread of carbon
abundances in these systems. Carbon, as well as oxygen, may have been
a key cooling agent inside the first galaxies \citep{dtrans}. One
extremely carbon-rich star (with $\mbox{[Fe/H]}\sim-3.5$) has already
been found in Segue\,1 \citep{norris10_seg}. 
This is consistent with
the predictions by \citet{dtrans}, but moreover, it adds to the
evidence that massive Pop\,III stars may have been the progenitors of
carbon-rich metal-poor stars. If stars with extremely low [C/Fe] and
[Fe/H] can be found, as has recently been done in the Milky Way \citep{caffau11}, it would
provide additional insights into early star formation in primitive
high-redshift halos.

The existence of such one-shot enrichment sites can be refuted with
future observations of UFD stars revealing the signatures of, e.g.,
s-process enrichment or [$\alpha$/Fe] abundance ratios systematically
lower than the halo value of $\mbox{[$\alpha$/Fe]}=0.35$ in these
galaxies. However, we would still have gained crucial empirical
constraints for the next generation of ab initio cosmological
simulations that will be able to resolve the fine structure of star
formation and feedback.  In the latter case, dwarf galaxy archaeology would
have indicated that negative feedback is not able to completely
suppress the ongoing formation of stars. Simulations could then adjust
their treatment of feedback accordingly. Alternatively, the absence of
any one-shot systems could simply indicate that we have not yet
discovered the true survivors of the first galaxies.

Simulations with extremely high-resolution that will study the
fine-grained turbulent mixing of metals on scales of a few AU will
soon become feasible. The dwarf galaxy archaeological comparison between UFDs
and early star forming halos is thus important for providing
constraints as well as consistency checks for state-of-the-art
simulations. The emerging field of dwarf galaxy archaeology, which closely
connects chemical abundances and galaxy formation models, promises a
more complete understanding of galaxy formation and evolution at the
end of the cosmic dark ages.

\acknowledgments We thank Ian Roederer and Lars Hernquist for useful
comments on an earlier version of the manuscript. A.~F. acknowledges
support of an earlier Clay Fellowship administered by the Smithsonian
Astrophysical Observatory. V.~B. acknowledges support from NSF grants
AST-0708795 and AST-1009928, as well as NASA ATFP grant NNX08AL43G.


\end{document}